\documentclass[reprint,
 superscriptaddress,
 amsmath,amssymb,
 aps,
]{revtex4-2}

\usepackage{graphicx}
\usepackage{dcolumn}
\usepackage{bm}
\usepackage{hyperref}
\hypersetup{hidelinks}
\usepackage{siunitx}
\usepackage[mathlines]{lineno}
\usepackage[dvipsnames]{xcolor}

\newcommand{\qi}{$Q_\mathrm{i}$ }

\newcommand{\nphoton}{$\left\langle n_\mathrm{ph}\right\rangle$ }

\newcommand{\detuning}{$\varepsilon$ }

\begin{document}

\title{3D integration of a hybrid quantum dot circuit-QED device for fast gate dispersive charge readout and coherent spin-photon coupling}
\author{Sébastien Granel}
\email{sebastien.granel2@cea.fr}
\affiliation{Univ. Grenoble Alpes, CEA, Grenoble INP, IRIG-Pheliqs, Grenoble, France.}
\author{Frédéric Gustavo}
\affiliation{Univ. Grenoble Alpes, CEA, Grenoble INP, IRIG-Pheliqs, Grenoble, France.}
\author{Jean-Luc Thomassin}
\affiliation{Univ. Grenoble Alpes, CEA, Grenoble INP, IRIG-Pheliqs, Grenoble, France.}
\author{Heimanu Niebojewski}
\affiliation{CEA-Leti, Univ. Grenoble Alpes, Grenoble, France.}
\author{Benoit Bertrand}
\affiliation{CEA-Leti, Univ. Grenoble Alpes, Grenoble, France.}
\author{Frédéric Berger}
\affiliation{CEA-Leti, Univ. Grenoble Alpes, Grenoble, France.}
\author{Alain Gueugnot}
\affiliation{CEA-Leti, Univ. Grenoble Alpes, Grenoble, France.}
\author{Chafik Mhamdi}
\affiliation{CEA-Leti, Univ. Grenoble Alpes, Grenoble, France.}
\author{Etienne Dumur}
\affiliation{Univ. Grenoble Alpes, CEA, Grenoble INP, IRIG-Pheliqs, Grenoble, France.}
\author{Romain Maurand}
\affiliation{Univ. Grenoble Alpes, CEA, Grenoble INP, IRIG-Pheliqs, Grenoble, France.}
\author{Simon Zihlmann}
\email{simon.zihlmann@cea.fr}
\affiliation{Univ. Grenoble Alpes, CEA, Grenoble INP, IRIG-Pheliqs, Grenoble, France.}

\date{\today}

\begin{abstract}
Hybrid circuit quantum electrodynamics (cQED) aims at coupling various quantum degrees of freedom, among which are spin and charge degrees of freedom in gate defined quantum dots, phonons or magnons... with quantized electromagnetic fields in superconducting microwave cavities to investigate fundamental physics questions or for quantum computation and simulation. However, low microwave losses, key for many hybrid cQED experiments, are challenging to achieve given the often exotic and/or complex material stacks (e.g. semiconducting material, ferromagnets, or piezoelectric materials) required to host the various quantum degrees of freedom. In this work, we present a 3D-integration process to overcome this challenge for semi-industrial silicon MOS spin qubits. The process is based on dense indium bump interconnects at a pitch of \SI{10}{\micro m} and superconducting thin films of Niobium Nitride (NbN). First, we report on DC and RF interconnect properties that demonstrate a high galvanic interconnection yield and internal quality factors above $10^5$ in the single photon regime for NbN resonators interrupted by a single indium bump interconnect. Eventually, we fabricate a 3D-integrated hybrid circuit quantum electrodynamics (cQED) device based on a semi-industrial MOS hole double quantum dot and a high impedance NbN resonator. For this device, we report a cavity internal quality factor above $10^4$ and demonstrate record sensitivity for gate-based dispersive readout of the charge degree of freedom with an $\mathrm{SNR}$ of $100$ in \SI{300}{\nano\second}. Finally, we demonstrate strong spin-photon coupling of $g_{s}/2\pi = 75\,MHz$, which highlights the viability of 3D-integration for quantum dot based hybrid spin cQED and opens to high-fidelity spin readout and microwave photon-based remote spin qubit entanglement.
\end{abstract}

\maketitle

\section{Introduction}
In cQED, the interaction between superconducting qubits and quantized electromagnetic fields in microwave resonators is engineered to allow for high-level qubit control, various coupling schemes, and for state readout with the aim of quantum information processing and simulation~\cite{2021_Blais}. With this promise, the spin qubit community started to explore the coupling of semiconducting quantum dots to superconducting microwave resonators~\cite{2012_Frey, 2012_Petersson, 2015_Viennot}. The use of high-impedance cavities led to demonstrations of strong charge-photon~\cite{2017_Stockklauser, 2017_Mi, 2024_DePalma, 2025_Janik} and spin-photon coupling~\cite{2018_Mi, 2018_Samkharadze, 2018_Landig, 2023_Yu, noirot_coherence_2026} as well as remote spin-spin interactions based on resonators~\cite{2019_Borjans, 2022_HarveyCollard, 2024_Dijkema}. However, the semiconducting environment hosting spin qubits is not necessarily well suited to host high quality superconducting circuits. For example, on Ge/Si-based heterostructure devices, one of the leading semiconducting platforms for spin qubits, quality factors above $10^4$ are obtained only when the heterostructure is removed at the resonator location~\cite{2025_Palma}. Similarly, integrating superconducting cavities with silicon MOS spin qubits generally leads to reduced quality factors of a few hundreds due to the complex material stack of MOS integration comprising various dielectrics, normal metals and spurious charge accumulations, which induce microwave losses ~\cite{2023_Yu}. These problems can be overcome by separating the microwave circuits from the semiconducting quantum dot devices in a 3D-integration, where two complementary chips are assembled together in a flip-chip processes. One chip hosts the semiconducting QD devices, whereas the second chip hosts the superconducting microwave circuit on a high-quality substrate. In doing so, fabrication and material constraints are reduced and both chips can be optimized independently. Proof-of-concept devices~\cite{2017_Foxen, Rosenberg2017, 2021_Conner} demonstrated the viability of this approach for superconducting qubits, while more recent reports focused on dense wiring and signal integrity~\cite{2022_Kosen, 2024_Kosena} in such 3D-integrations. The dispersive readout of a charge qubit in Si/SiGe quantum dots has recently been achieved in a 3D-integrated sample~\cite{Holman2021} using a process initially developed for superconducting qubits~\cite{Rosenberg2017}. However, strong charge- and spin-photon coupling in such a geometry remains to be demonstrated.

In this work, we present an approach to enhance the coherence of high-impedance superconducting niobium nitride cavities coupled to MOS spin qubit structures. It consists in leveraging thermomechanical bonding using indium bumps between a sapphire chip hosting the superconducting circuit and a MOS chip hosting a silicon hole spin qubit.
To access strong charge-photon and/or spin-photon coupling with gate-defined quantum dots, high-impedance superconducting cavities are essential as they ensure large zero-point voltage fluctuations ($V_\mathrm{zpf}$). Consequently, 3D integration should minimize parasitic capacitances to ground, which would reduce $V_\mathrm{zpf}$. To this end, we reduced the In bump size compared to conventional assemblies~\cite{Rosenberg2017, Holman2021}  using  \SI{5}{\micro m}$\times$\SI{5}{\micro m} bumps. We begin with the characterization of the flip-chip technology by evaluating the inter-chip galvanic connection yield and by probing the RF properties of the interconnects embedded in microwave resonators. We demonstrate the viability of the 3D-integration approach by reporting a quality factor of $Q_i \sim 10 000$ for a \SI{1.8}{\kilo\ohm} characteristic impedance NbN cavity coupled to a hole Si-MOS DQD. Using this device, we confirm that the 3D-integration has a minor impact on $V_\mathrm{zpf}$ by reporting a charge-photon coupling $g_\mathrm{c}/2\pi= $\SI{350}{\mega\hertz} associated with high quality charge dispersive readout. Eventually, at finite magnetic field we demonstrate state of the art strong spin-photon interaction. The work reported here provides a clear path to leverage cQED for spin qubits hosted in various semiconducting materials as well as the extension to other hybrid cQED platforms requiring high-impedance circuits.

\begin{figure*}
    \includegraphics[scale=0.88]{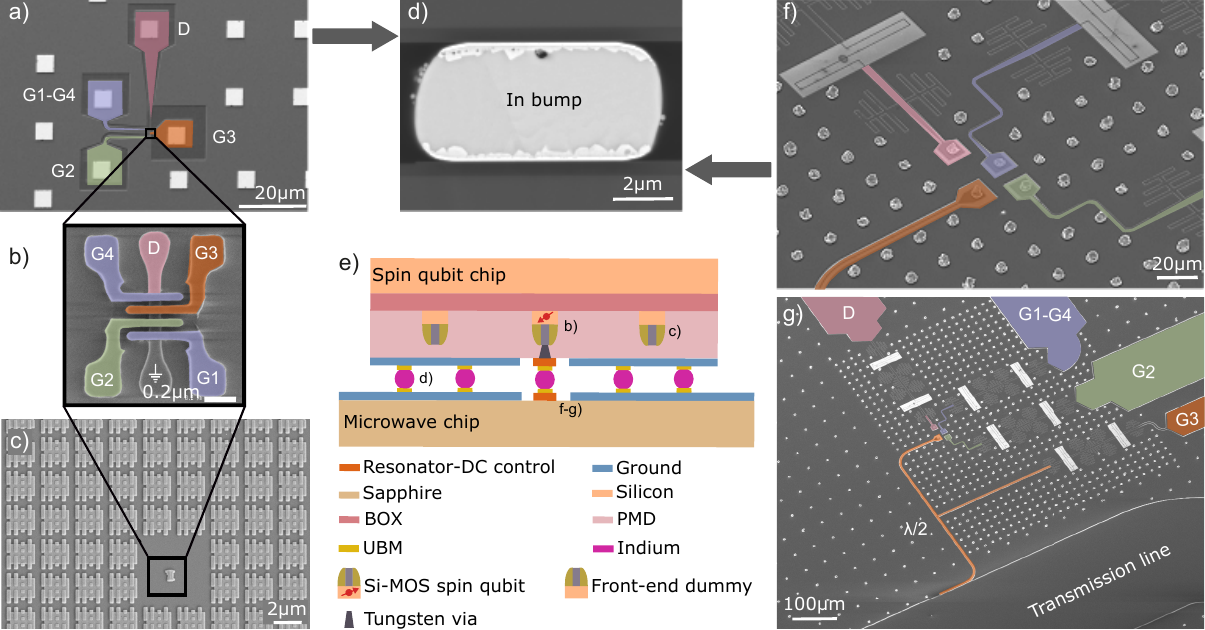}
    \caption{\textbf{Flip-chip architecture:} (\textbf{a}) 
    Top-view false-colour SEM image of a representative NbN circuitry patterned on the qubit-chip to connect the Si-MOS transistor to the control lines and the readout resonator. The areas where NbN is etched away appear in dark grey, while the UBMs appear as white squares. One open-end of the resonator shown in Figs.~\ref{fig1}-f) and g) is galvanically connected through an In bump interconnect to the gate $G3$ highlighted in orange. The source of the transistor is grounded at the MOS chip level while all remaining contacts, $D$ (pink), $G2$ (green) and $G1$ and $G4$ (both in purple) are galvanically connected through indium bumps to filtered DC control lines on the microwave chip as visible in Figs.~\ref{fig1}-f) and g). Note that $G1$ and $G4$ are shorted together on purpose at the MOS chip level to minimize the number of DC lines. (\textbf{b}) Top view false-colour SEM image of a representative four gate nanowire Si-MOS transistor. (\textbf{c}) Top view SEM image of the Si chip dummy structures buried under the PMD. (\textbf{d}) Cross-section SEM image of an Indium bump interconnect. (\textbf{e}) Cross-section schematics of the 3D-integrated hybrid spin cQED platform. The qubit hosts a silicon-based double quantum dot issued from a foundry-compatible metal–oxide–semiconductor fabrication process. The readout/control sapphire chip hosts the readout resonators, the readout feedline and the control lines for biasing the quantum dots. The two-chips are galvanically and thermo-mechanically bonded together with arrays of In-based bumps. (\textbf{f}) False-colour SEM image of the corresponding NBN circuit of (\textbf{d}) etched on the sapphire chip that comes to be flipped with the qubit chip. The areas where NbN is etched away appear in light grey. \SI{5}{\micro m}$\times$\SI{5}{\micro m} In-bumps are visible in bright white. First, we can notice arrays of bumps that aim at providing inter-chip mechanical support and ground-plane connectivity. Second, we distinguish "individual" bumps whose purpose is to connect resonator and DC bias-lines to the transistor. (\textbf{g}) False-colour SEM overview of the NbN microwave readout circuitry on the Sapphire chip. The DC-biased readout resonator (orange) is capacitively coupled to a \SI{50}{\ohm} feedline for RF transmission measurements and surrounded by DC-lines to control the DQD (pink, purple, green). Each of this control line is equipped with a $5^{th}$ order on-chip low-pass filter made of three fractal interdigital capacitors and two nanowire inductors.}
    \label{fig1}
\end{figure*}

\section{Indium interconnects and flip-chip assembly}
Figure~\ref{fig1}-e) illustrates the envisioned 3D integration between a microwave chip (bottom) and a Si-MOS semi-industrial spin qubit chip (top). The MW chip, visible in Fig.~\ref{fig1}-f) and -g) is based on a \SI{10}{nm} thick niobium nitride (NbN) film sputtered on a sapphire wafer. The superconducting circuit is defined by electron beam lithography and ICP dry etching, see methods for further information. The Si-MOS chip originates from a $300$\,mm semi-industrial fabrication process realizing nanowire based spin qubits similar to Ref. \cite{2016_Maurand, Piot2022, 2023_Yu, Bassi2025}, see Fig.~\ref{fig1}-b). The CMOS fabrication process of the qubit has been stopped at the pre-metal dielectric (PMD) planarization step prior to the first metallization layer of the back-end-of-line. Consequently, qubit devices and dummy structures (see Fig.~\ref{fig1}-c)) are buried in the PMD (SiO$_2$ here). Contacts from the surface to the device are made from tungsten vias. Similarly to the MW chip, the Si-MOS chip is covered by a \SI{10}{\nano\meter} thick NbN layer patterned with e-beam lithography and dry ICP etching realizing ground planes and the contact fan-out of the spin qubit device, see Fig.~\ref{fig1}-a) for illustration. Galvanic interconnects are realized by the use of an under bump metallization (UBM) on both chips consisting in a tri-layer of Ti/Pt/Au. Indium bumps are deposited on the MW chip following a process comprising optical lithography, UBM metalization, indium evaporation and lift-off. An indium bump lateral size of \SI{5}{\micro\meter} has been chosen with a minimal pitch of \SI{10}{\micro\meter} to reduce stray capacitances to ground. Figure~\ref{fig1}-d) shows a cross-section of an indium bump after assembly. The inter-chip distance is of the order of \SI{4}{\micro\meter} and the planarity, expressed as a relative tilt angle between the two chips, is below \SI{0.16}{\milli\radian}. 

\section{DC-characterization}
\begin{figure*}[htbp]
    \includegraphics[scale=1.04]{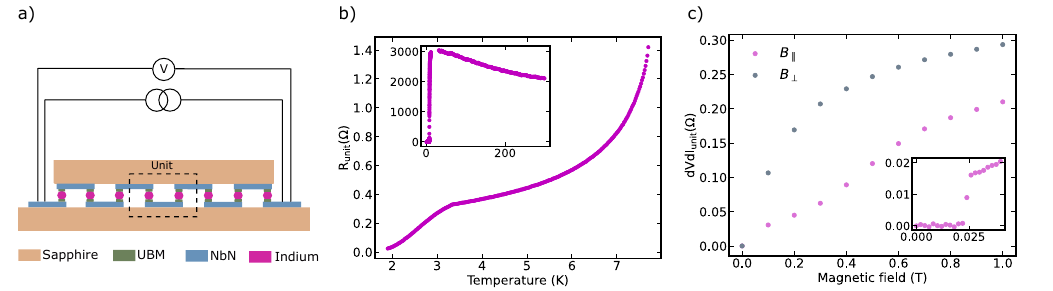}
    \caption{\textbf{DC characterization of the bump interconnects:} \textbf{(a)} Schematics of a daisy-chain containing four units (each containing two bump interconnects) used for DC characterization. \textbf{(b)} Resistance per unit of a 2000-unit daisy chain as a function of temperature. The largest fraction of the resistance originates from the sheet resistance of the disordered NbN layers~\cite{2021_Yu}, which turn superconducting at \SI{8.6}{K}. Below \SI{3.4}{K} the indium bumps become superconducting as well leading to a vanishing resistance with deceasing temperature. \textbf{(c)} Differential resistance per unit (dV/dI) of a 200 unit daisy chain as a function of magnetic field measured at \SI{10}{mK}. A zero-resistance state is observed up to the critical field of indium (\SI{24}{mT}).}
    \label{fig_DC}
\end{figure*}

We first electrically characterize the indium bump interconnects. Figure~\ref{fig_DC}~(a) shows a daisy chain containing four units, each of which is composed of two interconnects linking the two chips. A total of 44 daisy chains of 10 to 2000 units have been fabricated. From room temperature electrical characterization, we extract a single bump galvanic connection yield of \SI{99.98}{\percent}, see supplementary materials for details. Fig.~\ref{fig_DC}~(b) shows the resistance per unit of a 2000-unit daisy chain as a function of temperature. Two clear resistance drops are observed at \SI{8.6}{K} and \SI{3.4}{K} corresponding to the superconducting transition temperature of NbN and In respectively. At milli-Kelvin temperatures (well below $T_c$ of In), we find vanishing resistances ($R_\mathrm{unit}<$\SI{5}{\milli\ohm}), indicating that the interconnects turned entirely superconducting. Figure~\ref{fig_DC}~(c) presents the differential resistance of an interconnect unit \textit{versus} in-plane and out-of-plane magnetic fields. A field of \SI{24}{mT} is enough to suppress superconductivity in the interconnects associated with a resistance of one unit of approximately \SI{15}{m\ohm}, see Fig.~\ref{fig_DC}~(c) inset. With increasing magnetic field, Fig.~\ref{fig_DC}~(c), the unit resistance increases and tends to saturate around \SI{300}{m\ohm} (\SI{200}{m\ohm}) for \SI{1}{T} in-plane (out-of-plane) field. This resistance value is in agreement with Fig.~\ref{fig_DC}~(b) by taking $R_\mathrm{unit}$ at $T\approx3.2$~K just above the critical temperature of In. These resistance values are in-line with similar Indium bump-based integration reported in Ref.~\cite{Rosenberg2017}. In summary, a high galvanic interconnect yield has been reported for \SI{5}{\micro m} $\times$ \SI{5}{\micro m} In bump-based interconnects. Moreover, these interconnects show finite resistance values when operated in magnetic fields above the critical field of In.

\section{RF-characterization}
\begin{figure*}[htbp]
    \includegraphics[scale=0.4]{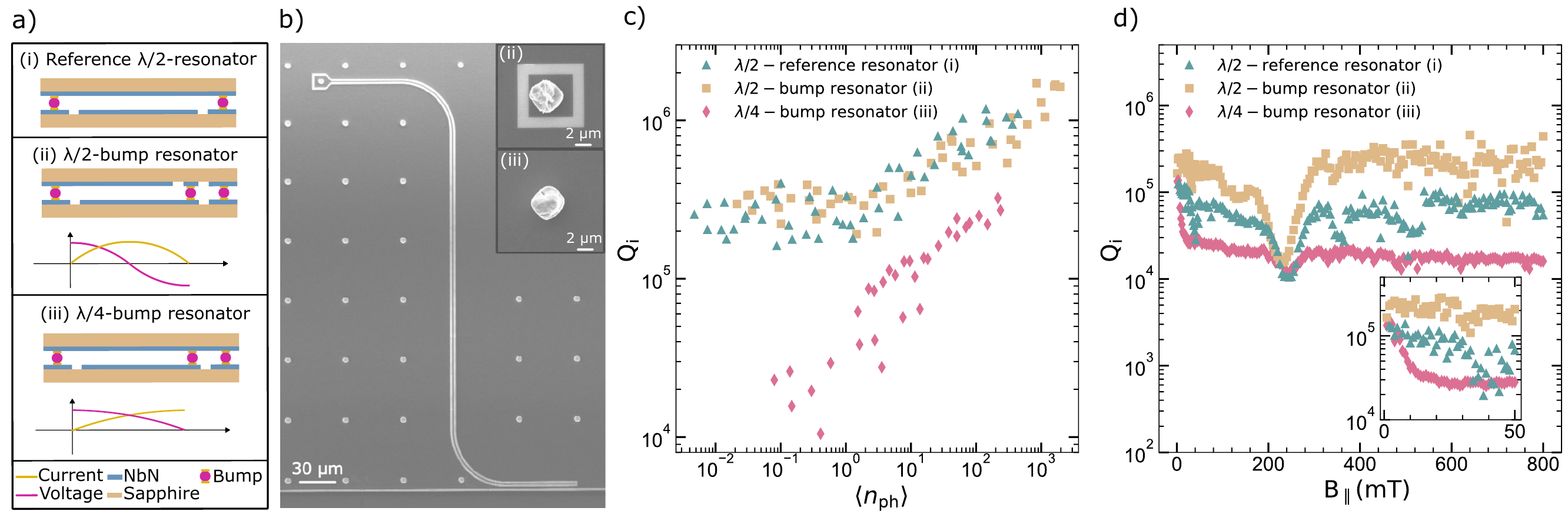}
    \caption{\textbf{RF-characterization of interconnects:} \textbf{(a)} Schematic of the three types of resonators used to assess high-frequency losses of In bump interconnects. A $\lambda/2$-resonator on the bottom chip serves as a reference (i), while a $\lambda/2$-resonator on the bottom chip with a bump at the voltage-antinode connecting to an isolated NbN-pad on the top chip (see also top inset in (b)) is predominantly probing dielectric losses of the interconnect (ii). A $\lambda/4$-resonator on the bottom chip with a bump at the voltage-node connecting to the ground plane on the top chip (see also bottom inset in (b)) is predominantly probing ohmic losses of the interconnect (iii). \textbf{b}) SEM image of a resonator on the bottom chip with a single bump at the end with the corresponding termination in the insets (top inset for type (ii) and bottom inset for type (iii)). \textbf{c}) Internal quality factor of type (i) - (iii) resonators as a function of average photon number. \textbf{d}) Internal quality factor of type (i) - (iii) resonators as a function of in-plane magnetic field (measured with \nphoton$\sim 50$). The pronounced dip around \SI{220}{mT} is due to coupling to magnetic impurities. 
    }
    \label{fig_RF}
\end{figure*}

To control spin qubit devices, interconnects will not only carry DC signals but also pulsed and microwave signals as for example through the interconnect between the gate $G3$ and the DC-biased $\lambda /2$ resonator of Fig.~\ref{fig1}. Therefore, interconnect  microwave losses (resistive dielectric) need to be quantified. To this end, we fabricated three types of coplanar waveguide (CPW) resonators: (i) a reference $\lambda/2$ resonator on the sapphire bottom chip. (ii) a $\lambda/2$ resonator, where one end is connected through a single interconnect to an NbN pad on the top chip to evaluate the influence of dielectric losses of the interconnect, see Fig.~\ref{fig_RF}~(a) top and (b) with top inset. (iii) a $\lambda/4$ resonator with a single interconnect shorting its central conductor to the ground plane of the top chip to evaluate the resistive loss of the interconnect, see Fig.~\ref{fig_RF}~(a) bottom and (b) with bottom inset. All CPW resonators have been designed to resonate between $5$ and \SI{7}{\giga\hertz} with a \SI{500}{\nano\meter}-wide central conductor and a \SI{2}{\micro\meter}-wide gap leading to a \SI{1.7}{\kilo\ohm} characteristic impedance. For each resonator type, we fabricated a series of resonators in a hanger-type geometry with varying coupling quality factor to reliable extract internal quality factors~\cite{2012_Megrant}. Such a geometry is visible for type (ii) in Fig.~\ref{fig_RF}~(b). Resonators of type (i) serve as reference and the internal quality factor \qi as a function of average photon number \nphoton of one such resonator is shown in Fig.~\ref{fig_RF}~(c). In the single photon limit \qi saturates around \num{2e5} probably limited by remaining resist residues~\cite{2024_Bahr} or oxide layers~\cite{2024_Bahr, 2025_Boettcher}. This \qi measurement establishes a reference to be compared to the \qi behavior of the other resonator types shown in Fig.~\ref{fig_RF}~(c). Resonators of type (ii) show a \qi behavior similar than the reference indicating that the additional dielectric loss at the interconnect is negligible at GHz frequencies. In contrast, resonators of type (iii) show a different behavior of \qi with \nphoton. The overall reduced \qi (approximately a factor of two to three) may be explained by additional resistive losses due to the In bump interconnect as previously observed~\cite{Rosenberg2017}. The absence of a saturation at \qi at very low photon number (\nphoton$<1$), however, needs further investigation.

Next, we assess the magnetic field resilience of the interconnect. Figure~\ref{fig_RF}~(d) shows the measurement of \qi (with \nphoton$ \sim 50$) as a function of in-plane magnetic field for all types of resonators. Type (i) and (ii) resonators exhibit \qi around \num{1e5} for fields up to \SI{0.8}{T}, except around $B\approx$\SI{220}{mT} where \qi is reduced. We associate this reduction in \qi with the resonant coupling to an ensemble of spin $1/2$, commonly observed for NbN~\cite{2021_Yu, 2026_Roy} and NbTiN~\cite{2016_Samkharadze, 2024_Bahr} resonators. For type (iii) resonators, the internal quality factor continuously decreases to \num{3e4} in the \SIrange{5}{25}{mT} range before stabilizing at \num{2e4}. We attribute this decrease to the transition of the interconnect from the superconducting to the normal state associated with an increase of the resistive loss. From an effective circuit model, we extract an effective high-frequency resistance of \SI{50}{m\ohm} in-line with the DC characterization, see supplementary material for further information.

\section{Coherent charge and spin photon coupling}
\begin{figure*}[htbp]
    \includegraphics[scale=0.35]{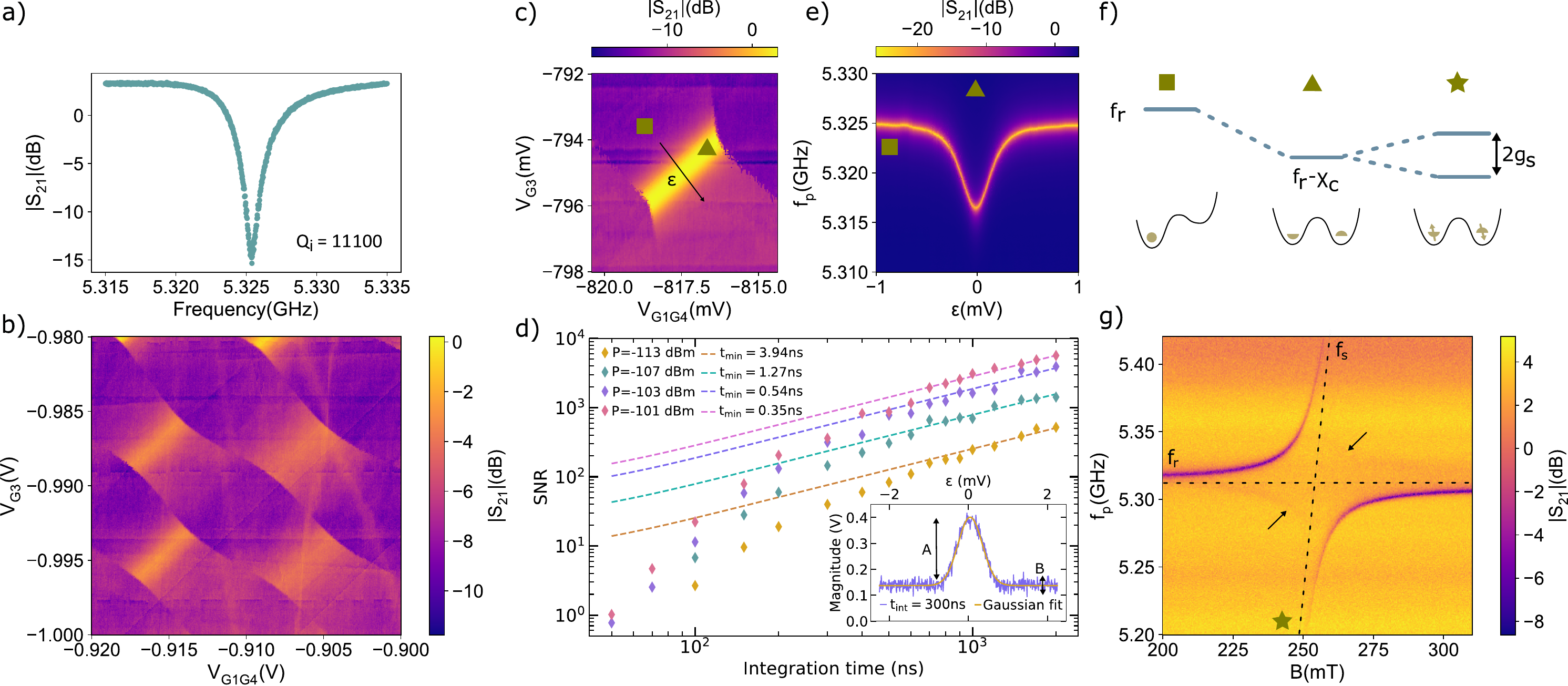}
    \caption{\textbf{Gate-based charge dispersive readout and strong spin-photon coupling in a 3D-integrated hybrid quantum dot cQED device}. \textbf{(a)} Transmission amplitude through the feedline as a function of probe frequency at zero magnetic field before forming QDs (e.g. in the Coulomb blockade region). In the single photon limit, at $\langle n_{ph} \rangle \approx 0.4$, we extract an internal quality factor $Q_{i}=11100$ and a coupling quality factor $Q_{c}=1500$. \textbf{(b)} Double quantum dot stability diagram in the many hole regime dispersively sensed with the microwave cavity. \textbf{(c)} Transmitted amplitude at \SI{5.325}{GHz} as a function of the voltages applied to the gates G3 and G1G4 at B=0 (note that gate G1 and G4 are shorted together at the device level). The energy detuning \detuning  is swept along the black arrow. \textbf{(d)} Charge readout $\mathrm{SNR}$ as a function of integration time. Four sets of data are shown, corresponding to input powers at the device level of \SI{-101}{dBm} (pink), \SI{-103}{dBm} (purple), \SI{-107}{dBm } (green) and \SI{-113}{dBm} (yellow). Each dataset is fitted well by a straight line for integration times long enough such that the cavity has reached its steady-state. The diverging behavior for small integration times is attributed to the transient regime, where the cavity is not completely populated. The inset shows the transmitted amplitude at $f_{r}$ around zero detuning at P\,=\,-103\,dBm at $t_{int}=$\,\SI{300}{\nano\second} integration time. \textbf{(e)} Transmission as a function of probe frequency $f_\mathrm{p}$ and \detuning. At large $|\varepsilon|$, the bare resonator is probed, whereas near \detuning=0, the DQD charge qubit dispersively interacts with the resonator, leading to a frequency shift of $\chi_c/2\pi$. \textbf{(f)} Schematic representation of the resonance frequency of the readout resonator. The bare resonator frequency (square) is dispersively shifted by $\chi_c/2\pi$ at \detuning=0 (triangle) and splits by $2g_{s}$ when it is resonant with the flopping mode spin qubit. \textbf{(g)} Transmission amplitude as a function of probe frequency $f_\mathrm{p}$ and magnetic field with \detuning=0 (triangle in (c)). An avoided crossing, the signature of strong spin–photon coupling, is observed when the spin transition frequency matches the resonator frequency. The magnetic field is applied in-plane of the sample with an angle of \SI{75}{\degree} with respect to the nanowire axis. Additional features inside the anti-crossing (marked by arrows) are due to a thermal population of the resonator as previously reported in Ref.~\cite{2023_Bonsen}.}
    \label{fig4}
\end{figure*}

To demonstrate the viability of the 3D-integration, we realize the flip-chip assembly corresponding to Fig.~\ref{fig1}. In the following, we configure the four-gate device to host a double quantum dot below gate  $G3$ and $G4$ by depleting the silicon channel under gate $G2$. Fig.~\ref{fig4}-a) shows the microwave transmission $\lvert S_{21} \rvert$ depending on the probe frequency when the device is in the Coulomb blockade regime. It reveals the resonance associated with the half-wavelength CPW resonator connected to the MOS device. For low probing powers (\nphoton$\leq 1$), we extract a bare resonance frequency $f_\mathrm{r}=\SI{5.325}{GHz}$ with an internal (coupling) cavity loss rate $\kappa_{i}/2\pi=\SI{540}{k\Hz}$ ($\kappa_{c}/2\pi=\SI{3.33}{MHz}$) corresponding to an internal (coupling) quality factor $Q_{i}=\num{1e4}$ ($Q_{c}=1500$). This internal quality factor is one order of magnitude lower than the test resonators presented in Fig.~\ref{fig_RF}-c). This reduction can be attributed to two main loss channels: (1) the remaining loss of the MOS-chip experienced by the resonator electric field at the level of the device, and (2) photon loss through the DC-control lines patterned in the vicinity of the resonator \cite{2022_HarveyCollard}. Further work is needed to determine and mitigate the remaining loss channel. Nevertheless, the \qi reported here competes favorably with the best 2D-integrated spin cQED devices \cite{2022_HarveyCollard, 2018_Mi} and places it in an ideal regime for fast spin and charge readout.

Probing the cavity at $f_\mathrm{r}$, Fig.~\ref{fig4}-b) presents the transmission $\lvert S_{21} \rvert$ as a function of the DQD voltages in the multi-hole regime. It reveals the dispersive interaction between the cavity and the charge qubits formed at each interdot transition of the DQD. To quantitatively access this interaction we focus on the interdot transition presented in Fig.~\ref{fig4}-c) on which we define the detuning axis $\epsilon$ controlling the energy difference between the two quantum dots. Measuring the hybrid system's response as a function of probe frequency and $\epsilon$, Fig.~\ref{fig4}-e) reveals a clear downshift of the cavity resonance at $\epsilon=0$. This is a clear signature of the electric-dipole interaction of the cavity with the DQD hole charge qubit of energy $\hbar \omega_\mathrm{c} = \sqrt{\epsilon^2 + 4t_\mathrm{c}^2}$ with $t_\mathrm{c}$ the interdot tunnel coupling. From the fit of the dispersive shift of the cavity depending on $\epsilon$ in Fig.~\ref{fig4}~(e), we extract a charge-photon coupling strength $g_\mathrm{c}/2\pi=\SI{350}{M\Hz}$ and tunnel coupling $t_{c}/h=\SI{14.5}{G\Hz}$, see supplementary information for the extraction of the gate lever arms necessary for the fitting procedure. The measured charge-photon coupling agrees with the theoretically expected $g_\mathrm{c}^{\mathrm{th}} = \alpha_{G3} e V_\mathrm{zpf}/(2h)$ with $V_\mathrm{zpf}$ the zero-point voltage fluctuation of a \SI{1.8}{k\ohm} impedance resonator, $\alpha_{G3}$ the lever arm of gate G3, $e$ the electron charge and $h$ the Planck constant. The measured charge-photon coupling is $ 5-10$\% lower than the theoretically expected $g_\mathrm{c}^{\mathrm{th}} = \alpha_{G3} e V_\mathrm{zpf}/(2h)$ with $V_\mathrm{zpf}$ the zero-point voltage fluctuation of a \SI{1.8}{k\ohm} impedance resonator, $\alpha_{G3}$ the lever arm of gate G3, $e$ the electron charge and $h$ the Planck constant. We attribute this decrease to the added stray capacitance originating from the bump interconnect between the resonator and the MOS chip. More precisely, to realize the interconnect the central conductor of the CPW resonator has been enlarged from \SI{800}{\nano\meter} to \SI{8}{\micro\meter}, which adds an estimated capacitance of $\sim$\SI{2.5}{\femto\farad} that reduces $V_\mathrm{zpf}$, see supplementary information for details.

We now evaluate the sensitivity of the cavity to changes of the DQD admittance as a hole is tunneling from one dot to the other. To this end, we sweep over the detuning axis and perform a pulse readout measurement at the bare resonance frequency $f_\mathrm{r}$ for each detuning value, see inset of Fig.~\ref{fig4}-d) for such a measurement with an integration time of \SI{300}{ns} at a readout power P=\SI{-103}{dBm} at the device level. The transmitted microwave signal is amplified using a traveling-wave parametric amplifier (TWPA) in the measurement chain to further improve the signal-to-noise-ratio. Following Ref.~\cite{2019_Zheng}, the power $\mathrm{SNR}$ is defined as $\mathrm{SNR}=(A/B)^{2}$. The signal $A$ is obtained from a gaussian fit of the difference between the transmitted amplitude at the interdot transition (triangle) and the Coulomb blockaded region (square). The noise $B$ is the standard deviation of the amplitude measured in the Coulomb blockaded region. Fig.~\ref{fig4}-d) shows the $\mathrm{SNR}$ measured for integration times spanning from \SI{50}{\nano\second} to \SI{2}{\micro\second} for four different probe powers. In the experiment, the acquisition window is aligned with the beginning of the \SI{2}{\micro \second} long readout pulse. 
Due to the characteristic filling time of the resonator(1/$\kappa$\,=\,\SI{33}{ns}), we distinguish two regimes in the $\mathrm{SNR}$ behavior with respect to the integration time. For integration times long enough such that most of the signal is acquired while the cavity is in its steady-state, the $\mathrm{SNR}$ scales linearly with respect to the integration time, similarly to the continuous wave measurements of refs~\cite{2019_Zheng,Ibberson_2021}. For each power, a minimum integration time $t_{min}$ corresponding to an SNR of unity can be extrapolated from the slope of a linear fit $\mathrm{SNR}(t_\mathrm{int}>>\kappa^{-1})=t_\mathrm{int}/t_\mathrm{min}$. At $-103$\,dBm, we extract $t_{min}\,=\,$\SI{0.54}{ns}, which is close to be 10 times shorter than $t_{min}$ extrapolated at $-113$\,dBm, as expected due to the $10$\,dB difference in power. At short integration times $t_\mathrm{int} \approx 1/\kappa$, the data diverge from the linear fit with a $\mathrm{SNR}$ dominated by the transient dynamics of the cavity.
Despite the limiting transient dynamics of the resonator, for $P=-103$ dBm, we report $\mathrm{SNR}\approx 100$ at $\tau_{int}=$\SI{300}{ns} and $\mathrm{SNR}\approx 10^3$ at $\tau_{int}=$\SI{1}{\micro s}. These $\mathrm{SNR}$ values are reached with integration times almost an order of magnitude shorter than previously reported~\cite{2019_Zheng, Ibberson_2021}. This demonstrates the ability of the 3D integrated hybrid cQED platform to perform fast and highly sensitive charge sensing.

Eventually we finalized our study by investigating the ability of the 3D-integrated device to demonstrate a coherent spin-photon coupling. Due to the intrinsic spin-orbit interaction at play in the valence band of silicon, spin transitions of holes inherit part of the charge electric dipole allowing spin-photon coupling~\cite{2023_Michal,Fang2023}. Following ref~\cite{2018_Mi, 2018_Samkharadze, 2023_Yu, noirot_coherence_2026} we investigate the spin-photon interaction at $\epsilon=0$, where the electric dipole of the hole charge in the DQD is maximized. By applying an external magnetic field $B \approx 250$\,mT, we bring the Zeeman spin splitting energy $g\mu_\mathrm{B}B$, with $g$ the effective g-factor and $\mu_B$ the Bohr magneton, in resonance with the frequency of the MW cavity. Figure~\ref{fig4}-g) presents the microwave transmission depending on the probe frequency $f_\mathrm{p}$ and the external in-plane magnetic field $B$. At the resonance condition $g\mu_\mathrm{B}B/h = f_\mathrm{r}$, highlighted by a star symbol in Fig.~\ref{fig4} f) and f), spin-photon coupling is revealed as an avoided crossing splitting the cavity response into two branches separated by the vacuum Rabi mode splitting $2g_{s}/2\pi=\SI{150}{M\Hz}$ with $g_{s}/2\pi=\SI{75}{M\Hz}$ the spin-photon coupling clearly exceeding the spin and photon linewidth.

\section{Discussion and conclusion}
Thermomechanical bonding using indium bump interconnects allows to realize high-quality superconducting microwave resonators that find applications in cQED experiments probing charge and spin degrees of freedom in gate defined semiconducting quantum dots. The small, but finite, resistance associated with a single In bump and its under bump metallization, does not impact the quality factor of a resonator as long as the interconnect is placed at a voltage anti-node. Lower resistive losses (currently dominated by the normal metal UBM), may be achieved by using superconducting materials such as TiN~\cite{2017_Foxen} or NbN~\cite{2022_Kosen}. In addition, indium interconnects could be replaced by direct bonding of Nb pillars~\cite{2024_Renaud} to ensure reduced losses at elevated temperatures and magnetic fields.

The high-quality of the In bump interconnects has been characterized by a high-yield of galvanic connections. The small size of the In bump interconnects of only \SI{5}{\micro m}\,$\times$\,\SI{5}{\micro m} is crucial in order to preserve the high-impedance of the microwave resonator as the parasitic capacitance (\SI{2.5}{fF}) added is kept low. As the stray capacitance scales with the area of the In bump interconnect, a further reduction in bump size will allow for even higher impedance circuits and hence larger light-matter couplings.

The successful 3D integration of an hybrid cQED device, comprising a \SI{1.8}{\kilo\ohm} resonator coupled to a hole Si-MOS double quantum dot, demonstrates the platform's potential. Beyond silicon MOS qubits, this 3D integration scheme is adaptable to other semiconducting materials (Ge/SiGe or Si/SiGe heterostructures or III-V quantum dots) and even possibly other hybrids systems (e.g. magnon or phonon). With the MOS device, we achieve a charge-photon coupling of $g_\mathrm{c}/2\pi=$\SI{350}{\mega\hertz}. This results in charge dispersive readout with a signal-to-noise ratio of $100$ in \SI{300}{\nano\second}, surpassing prior implementations laying the basis for rapid high-fidelity spin readout either through parity measurements~\cite{2019_West, 2019_Urdampilleta, 2019_Zheng} or direct dispersive spin readout~\cite{2025_Chessari}. Longitudinal readout schemes \cite{2023_Corrigan, 2025_Chessari, 2025_Champain, 2025_Harpt, 2025_Jarjat} could further enhance the fidelity of spin readout at short integration times.

In summary, we have demonstrated that semi-industrial flip-chip technology based on In bump interconnects - originally developed for infrared applications- enables the creation of high-quality high-impedance microwave environments for hybrid cQED systems. The successful realization of a high-quality spin cQED architecture, evidenced by the observation of strong-spin photon coupling ($g_\mathrm{s}/2\pi=$\SI{75}{\mega\hertz}), underscores the full potential of this approach. Futur work will focus on multi-qubit entanglement via microwave photons and the exploration of spin readout schemes. We anticipate that these proof-of-principle experiments will serve as a foundational step toward spin cQED architecture capable of supporting large-scale quantum computing and simulation with spin qubits.

\section{Methods}
Two types of samples have been fabricated in this study: (1) flip-chip assemblies containing two sapphire chips with patterned NbN circuits (DC and microwave components) on both of them, see Ref.~\cite{2021_Yu} for details and (2) flip-chip assemblies containing a NbN microwave circuit on a sapphire chip (bottom) and a silicon chip hosting silicon nanowire MOS transistors with a NbN routing circuit that replaces the first metallic interconnect layer of the back-end-of-line, see Ref.~\cite{2023_Yu} for details.

The fabrication of the flip-chip assemblies starts with the deposition of a \SI{10}{nm} thick NbN layer on each chip/wafer by DC magnetron sputtering, see Ref.~\cite{2021_Yu} for details. E-beam lithography and dry etching based on fluorine chemistry are then used to pattern the NbN layers on both chips. UBM pads of \SI{5}{\micro m} by \SI{5}{\micro m} or \SI{7}{\micro m} by \SI{7}{\micro m} with a minimal pitch of \SI{15}{\micro m} are patterned with UV-lithography and subsequent evaporation of \SI{50}{nm} Ti/\SI{100}{nm} Pt / \SI{50}{nm} Au. Thermal evaporation is used to deposit \SI{5}{\micro m} of In on the bottom chip using the same photo resist as for the UBM. After dicing the wafer/chips into their final size and after careful cleaning with solvents, the two chips are assembled on a Pick\&Place SET FC300 machine, see Refs.~\cite{2019_MAILLIART, 2023_Feautrier} for details. Excellent lateral alignment and planarity is achieved by a reflow of the indium interconnects after assembly.

Microwave characterization of the final flip-chip assemblies is performed in a dry dilution refrigerator (base temperature of \SI{10}{mK}) equipped with a 3D vector magnet (6–1–1 T) and connected to a standard microwave setup, see supplementary for details.

\begin{acknowledgments}
This work is supported by the program QuanTEdu-France n°ANR-22-CMAS-0001 France 2030.
This research has been supported by the European Union’s Horizon 2020 research and innovation programme under grant agreements No. 951852 (QLSI project), No. 810504 (ERC project QuCube), No. 759388 (ERC project LONGSPIN), No. 101174557 (QLSI2) and by the National strategy France 2030 under the project PEPR PRESQUILE - ANR-22-PETQ-0002 and PEPR MiraclQ ANR-23-PETQ-0003. E. Dumur acknowledges support from ANR through the HARDWAVE project (ANR-23-CE47-0010)). S. Zihlmann acknowledges support by the spin-photon PEPR chair. 
The authors dedicate this paper to Mathis Fragnol.
\end{acknowledgments}

\section*{Author contributions}
S.G. and F.G. fabricated the NbN circuitry for DC and RF characterization with help from J.L.T and C.M. S.G. fabricated the NbN circuitry of the hybrid device with help from F.G. and J.L.T. F.B. made the flip-chip assembly. S.G. performed the measurements with inputs from S.Z. and R.M. S.G. analysed the data with inputs from S.Z., R.M. and E.D. S.G., S.Z. and R.M. co-wrote the manuscript with inputs from all the authors. H.N and B.B. were responsible for the front-end fabrication of the hybrid device. S.Z. and R.M. initiated the project.

\nocite{*}

\clearpage
\newpage
\section{Supplementary information}

\subsection{Wiring scheme}
\begin{figure*}[htbp]
     \includegraphics[scale=0.65]{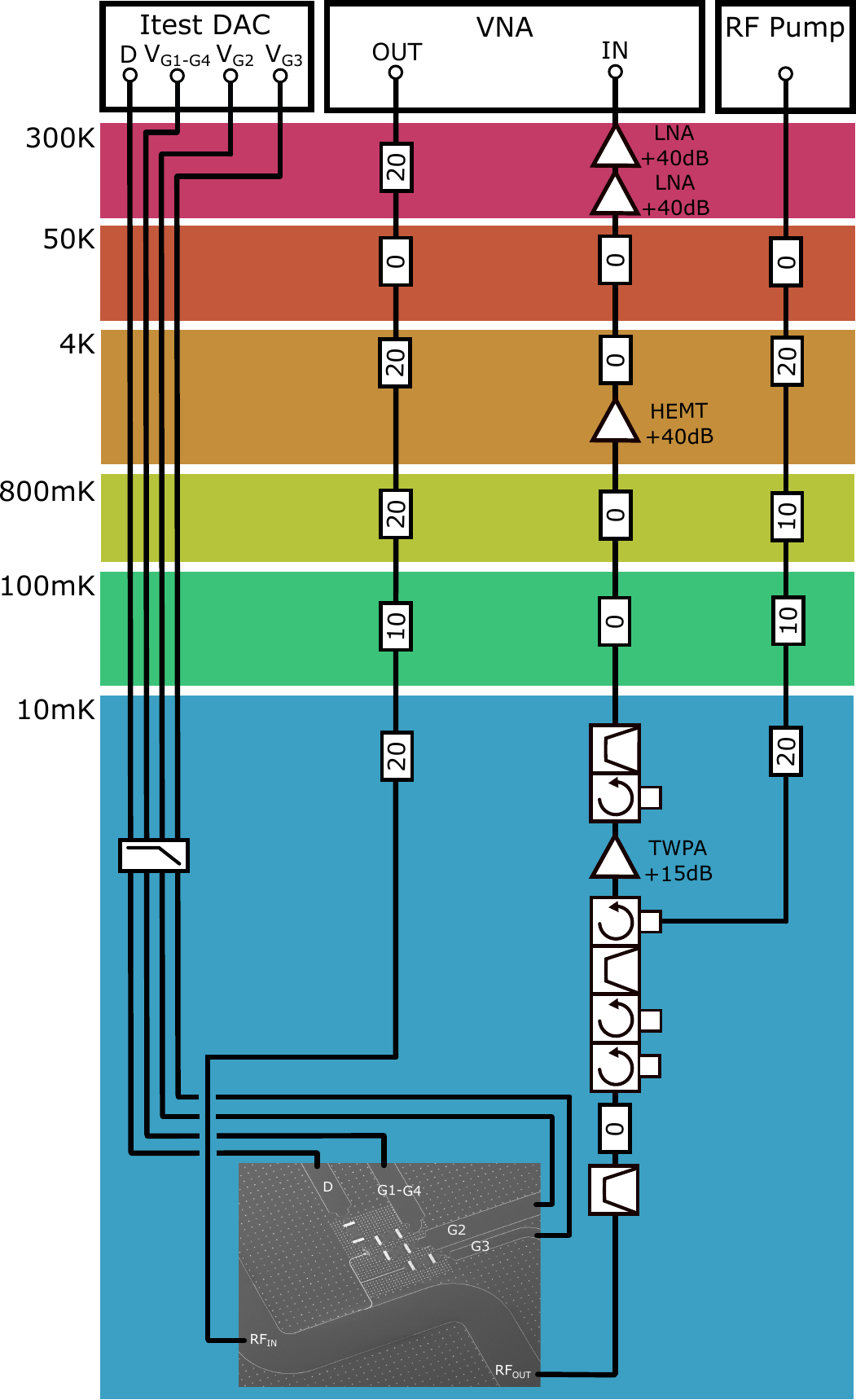}
     \caption{\textbf{Measurement setup:} Schematic of the measurement setup for a DQD hybrid architecture. The TWPA was only used for charge readout SNR measurements discussed in Fig.4c of the main text. For the RF characterization of interconnects, resonator's response was probed only using the RF lines connected to the VNA in the schematics without the TWPA. }
     \label{fig:s1}
\end{figure*}

All measurements of the resonators and hybrid structure are performed at $T=\SI{8}{\milli\kelvin}$ in a dilution refrigerator equipped with a 3D vector magnet (6-1-1 T) and connected to a standard microwave setup, see Fig.~\ref{fig:s1}. The input line of the resonator has -90 dB discrete attenuation and the readout line is equipped with a cryogenic low noise HEMT amplifier at 4K and two amplifiers at room temperature. Furthermore, the readout line is equipped with a TWPA (BDWCM model from Arctic) for the characterization of the hybrid cQED device. Since the noise is limited by the HEMT, the TWPA was turned on for the charge readout SNR measurements to reduce the noise temperature. The superconducting resonators are measured in transmission using a two-port network analyzer (VNA), MF5180 from Copper mountain and/or ZNB 8 from Rohde $\&$ Schwarz. The time domain measurements for charge readout SNR are performed with a QBlox cluster. The DC gate voltages are supplied by a BE2231 card in a Bilt rack from Itest and are low pass filtered at mixing chamber temperature (multi stage LC and RC filters).
\subsection{Bump interconnects success rate}
\begin{figure}[htbp]
     \includegraphics[scale=0.5]{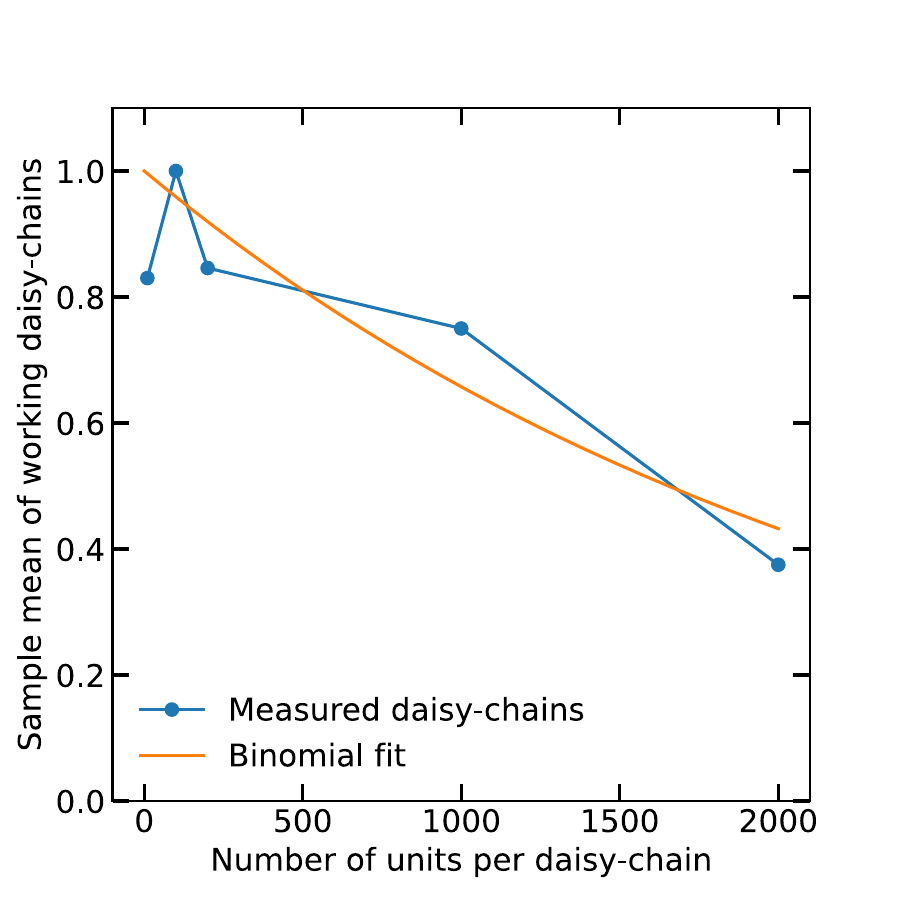}
     \caption{\textbf{Normalized number of working daisy-chains of different lengths} }
     \label{fig:s2}
\end{figure}
Extraction of the success rate of bump interconnects relies on room temperature four-probe electrical measurement of daisy-chains of bumps.
In the following, a daisy-chain is defined as ``working" if it satisfies two conditions. First, if the chain is not shorted to ground, and second, if the daisy-chain resistance measured at room temperature matches the product of the NbN sheet resistance by the number of NbN squares of the chain.
In order to estimate the probability of success rate $p$ that one bump provides inter-chip galvanic connection, we model the number of working bumps in an N-units (one unit contains two bumps) long daisy-chain as a random variable $X$ following a binomial distribution $X\rightsquigarrow B(2N,p)$. 
As a result, the probability that a daisy-chain of N units will work is written $P(X=2N)=p^{2N}$.
From a total of 44 daisy-chains of 10 to 2000 units that have been tested we extract the mean number of working daisy-chains with respect to the number of units they are made of. A fit of this dependence with the binomial model enables us to estimate $p=99.98 \%$, see Fig.~\ref{fig:s2}.

\subsection{High-impedance resilience to flip-chip integration}
\begin{figure*}[htbp]
     \includegraphics[scale=0.7, angle=-90]{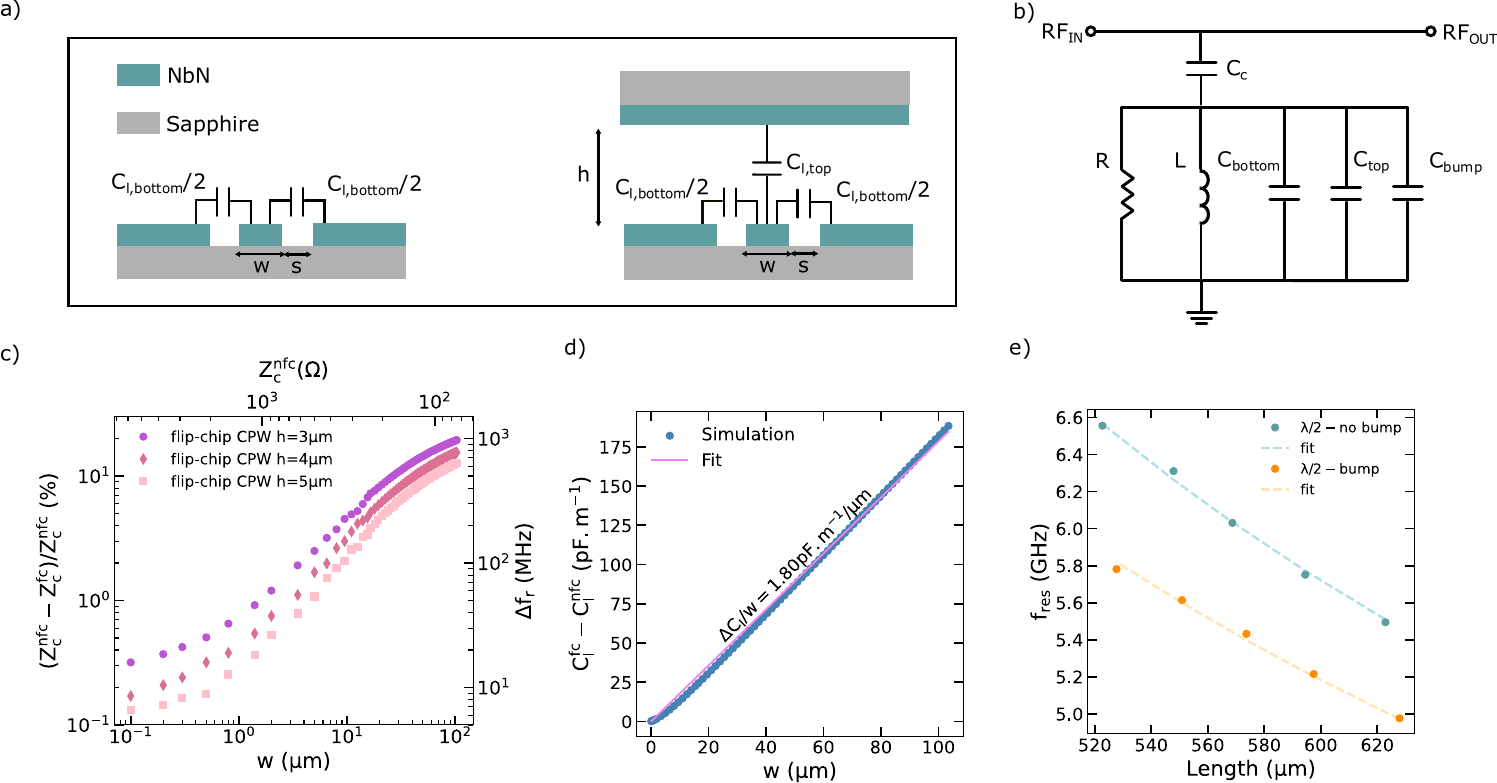}
     \caption{\textbf{High-impedance compatible flip-chip integration:} \textbf{(a)} Schematics of the cross-section of a CPW geometry facing air (on the left) in a non-flip-chip assembly (nfc) or facing NbN (on the right) in a flip-chip assembly (fc). The top NbN ground plane provides a lineic capacitance ($C_{l}^{top}$). \textbf{(b)} General lumped-element circuit model of a resonator in a flip-chip assembly. For a resonator patterned on the bottom chip, $C_{bottom}$ stands for the capacitance to the ground of the bottom-chip, $C_{top}$ illustrates the capacitance of the resonator CPW geometry to the top ground chip and $C_{bump}$ stands for the capacitance picked by the resonator due to the enlargement of the CPW geometry at the bump-level. \textbf{(c)} Simulated change in $Z_{c}$ with respect to the CPW central conductor width when a CPW faces NbN for different inter-chip spacing compared to a CPW facing air (all the simulations are made with $L_{kin}=\SI{200}{\pico \henry}/\square$ NbN thin film on both chips). The top axis outlines the characteristic impedance exhibited by a given CPW geometry facing air with a central conductor width corresponding to the value reported on the bottom axis. The right axis stresses how much the resonance frequency of a 5GHz targeted non-flip-chip resonator shifts due to $Z_c$ change when the CPW faces NbN in a flip-chip assembly. \textbf{(d)} Simulated gain in the lineic capacitance of the CPW that originates from a top NbN ground plane placed \SI{4}{\micro\meter} above the CPW geometry in a flip-chip assembly. The simulation investigates CPW geometries whose central conductor width spans from 100 nm to 100 µm. \textbf{(e)} Measured frequencies of a set of type (i) (blue) and type (ii) (orange) resonators versus physical length. For a same lenght, the  resonance frequency of type (ii) resonators is downshifted compared to type (i) resonators because of the stray capacitance that originates from the enlargement of the central conductor at the bump level. }
     \label{fig:s3}
\end{figure*}

In order to reach a strong charge-photon coupling and/or spin-photon coupling, it is crucial to maintain high-impedance in the 3D assembly. In this scope, two aspects of the 3D integration have been identified as possible sources of impedance decrease.\\
On the one hand, placing an NbN-covered chip above the CPW geometry adds a stray capacitance $C_{l, top}$ of the central conductor to the ground, see Fig.~\ref{fig:s3}{a}. Simulations performed with Sonnet were used to evaluate how much this added capacitance decreases the CPW characteristic impedance for different inter-chip spacings and center conductor widths (the gap is fixed at 2 µm and NbN sheet inductance at \SI{200}{pH/}$\square$). The results in Fig.~\ref{fig:s3}{c} show that while low \SI{70}{\ohm} impedance CPW associated with \SI{100}{\micro\meter} wide center conductor exhibit 20$\%$ decrease of the impedance, \SI{2}{\kilo\ohm} high-impedance CPW associated with \SI{800}{\nano\meter} wide center conductor exhibit less than 1$\%$ impedance decrease. The stray lineic capacitance that originates from the top ground plane is estimated to be $C_{l, top}=\Delta C_{l}=\SI{1.80}{\pico\farad}$.m$^{-1}$ per µm added to the width of the central conductor, see Fig.~\ref{fig:s3}{d}. As a result, the top capacitance is not detrimental for enabling k$\Omega$-high-impedance.
\\ 
On the other hand, placing a bump at one end of the resonator has been identified as adding an extra-capacitance of the CPW to the ground. Indeed, enlargement of the CPW central conductor from \SI{500}{\nano\meter}-\SI{800}{\nano\meter} to \SI{8}{\micro\meter} is required to interconnect the resonator, resulting in an increase of the lineic capacitance to ground at the level of the enlarged segment. 
In order to estimate the capacitance that originates from this enlargement of the CPW geometry, we measured resonance frequencies of a set of type (i) and type (ii) resonators discussed in the main text. A fit of the frequency dependence with respect to their length is reported in Fig.~\ref{fig:s3}{e}. Resonators interrupted with a bump downshift of approximately $\Delta f_{r}=$\SI{500}{\mega\hertz} compared to type (i) reference resonators of the same length. This shift of the resonance frequency then allows us to extract the effective bump capacitance $C_{bump}$ from a fit of $f_{r}$ with:
\begin{equation}
f_{r}=\frac{1}{2\pi}\frac{1}{\sqrt{\frac{l}{2}L_{l}({\frac{2l}{\pi^2}C_{l}+C_{bump}))}}}
\end{equation}
where $l$ is the resonator's length and  $L_{l}=\SI{260}{\micro\henry}.$m$^{-1}$ and $C_{l}=C_{l, bottom}+C_{l, top}=\SI{87}{\pico\farad}.$m$^{-1}$ are the lineic inductances and capacitances of a $w=\SI{500}{\nano\meter}$ and $s=\SI{2}{\micro\meter}$ flip-chip CPW patterned on a $L_{kin}=\SI{130} {\pico\henry}/\square$ NbN thin film. The bump capacitance extracted therefore writes $C_{bump}=$2.31 $\pm$ 0.2 fF. It should be noted that this bump capacitance constitutes almost $25\%$ of the total capacitance of a $l$=\SI{600}{\micro\meter} long $w$=\SI{500}{\nano\meter} CPW resonator, resulting in $10-15 \%$ decrease in $Z_{c}$. As a consequence, interconnection with $\SI{5}{\micro\meter}$ small bumps is not detrimental for maintaining high-impedance with $Z_{c}\approx\SI{2}{\kilo \ohm}$ CPW resonators. 
\\
A summary of the different capacitive contributions from flip-chip integration is drawn in Fig.~\ref{fig:s3}{b}, where the lumped-element $RLC_{bottom}$ model of a non-flip-chip CPW is refined with the parallel capacitances $C_{top}$ and $C_{bump}$.

\subsection{Power dependence and magnetic field dependence of the resonators' internal quality factors.}
\begin{figure*}[htbp]
    \includegraphics[scale=0.8, angle=0]{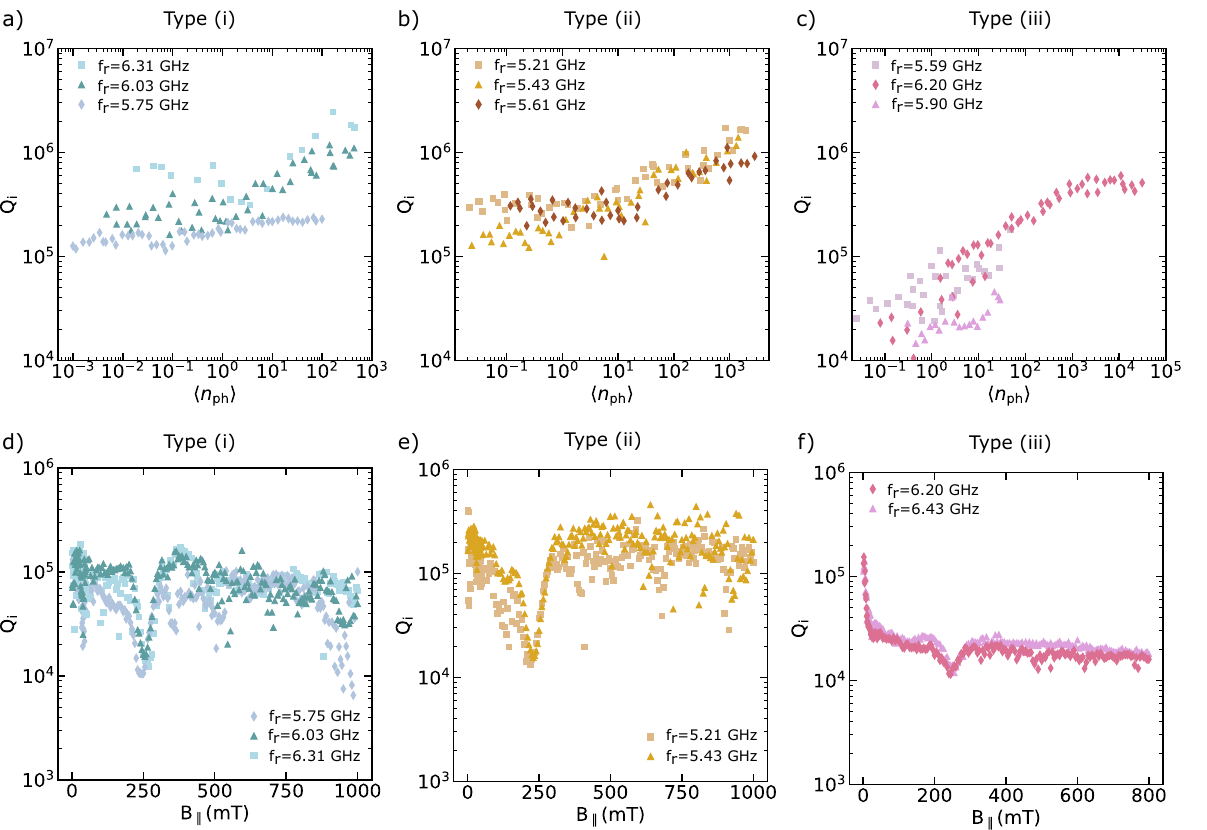}
     \caption{\textbf{Power dependence and magnetic field dependence of the resonators internal quality factor:} 
     \textbf{(a)},\textbf{(b)},\textbf{(c)}: 
    Full dataset of the internal quality factors as a function of the average photon number in the resonators.\textbf{(a)} $Q_{i}$ as a function of $\langle n_{ph} \rangle $ for type (i) reference resonators with different fundamental resonance frequencies. \textbf{(b)} $Q_{i}$ as a function of $\langle n_{ph} \rangle $ for type (ii) resonators. \textbf{(c)} $Q_{i}$ as a function of $\langle n_{ph} \rangle $ for type (iii) resonators. 
     \textbf{(d)},\textbf{(e)},\textbf{(f)}: Full dataset of the internal quality factors as function of the in-plane magnetic field magnitude for a fixed number of $\approx$ 50 photons. \textbf{(d)} $Q_{i}$ as a function of $B_{\parallel}$ for type (i) reference resonators with different fundamental resonance frequencies.\textbf{(e)} $Q_{i}$ as a function of $B_{\parallel}$ for type (ii) resonators.\textbf{(f)} $Q_{i}$ as a function of $B_{\parallel}$ for type (iii) resonators.}
    \label{fig:s4}
\end{figure*}

In Fig.~\ref{fig:s4}, we show the dataset of the internal quality factor as a function of the average number of photons and the magnetic field magnitude measured for several resonators of each type. For type (i) resonators, we notice in Fig.~\ref{fig:s4}{a} that $Q_{i}$ of all the resonators saturate in the single photon limit. However, the $Q_{i}$ value in the single-photon regime is not exactly the same for all. While $f_{r}=\SI{5.75}{\giga\hertz}$ and $f_{r}=\SI{6.03}{\giga\hertz}$ resonators saturate around $Q_{i} \approx 2 \times 10^{5}$, $f_{r}=\SI{6.31}{\giga\hertz}$ resonator exhibit scattered data around $Q_{i} \approx 5 \times 10^{5}$. Furthermore, while the $f_{r}=\SI{6.31}{\giga\hertz}$ and $f_{r}=\SI{6.03}{\giga\hertz}$ resonators both exhibit increasing $Q_{i}$ with respect to increasing $\langle n_{ph} \rangle $ for $\langle n_{ph} \rangle >1 $, the $Q_{i}$ of the $f_{r}=\SI{5.75}{\giga\hertz}$ resonator increases much less with respect on the mean number of photons for $\langle n_{ph} \rangle >1 $.\\
In Fig.~\ref{fig:s4}{b}, the power dependence of all the measured type (ii) resonators' quality factors is similar. \\
In Fig.~\ref{fig:s4}{c}, $Q_{i}$ of all the measured type (iii) resonators exhibit $Q_{i}$ of the same order of magnitude around $3 \times 10^{4}$ in the single photon limit. From measurement of the $f_{r}=\SI{6.20}{\giga\hertz}$ resonator, we notice that $Q_{i}$ increases with increasing $\langle n_{ph} \rangle$ up to $\langle n_{ph} \rangle \approx 10^{3}$. For $\langle n_{ph} \rangle > 10^{3}$, $Q_{i}$ saturates around $5 \times 10^{5}$. Such saturation in $Q_{i}$ may be attributed to the creation of quasi-particles at the In bump level \cite{sun_quasiparticle_2026}. \\
Fig.~\ref{fig:s4}{d} and Fig.~\ref{fig:s4}{e} show measurements of the magnetic field dependence of type (i) and type (ii) resonators' quality factors respectively. We notice that all the measured resonators exhibit the same behavior. They all have $Q_{i}$ around $1 \times 10^{5}$ for fields up to $\SI{0.8}{\milli\tesla}$, except around $B=\SI{220}{\milli\tesla}$ where $Q_{i}$ is reduced to $1 \times 10^{4}$ due to the resonant coupling to an ensemble of spin 1/2. \\
The magnetic field dependence of the type (iii) resonator's quality factors is shown in Fig.~\ref{fig:s4}{f}. For all measured resonators, we observe a continuous decrease in $Q_i$ from $\SI{5}{\milli\tesla}$ to $\SI{25}{\milli\tesla}$ before stabilizing at $2 \times 10^{4}$. We also notice a dip in $Q_i$ around $B=\SI{220}{\milli\tesla}$ due to the resonant coupling  of the cavity to an ensemble of spin 1/2. 

\subsection{RF resistance of a bump interconnect}
\begin{figure*}[htbp]
     \includegraphics[scale=0.8]{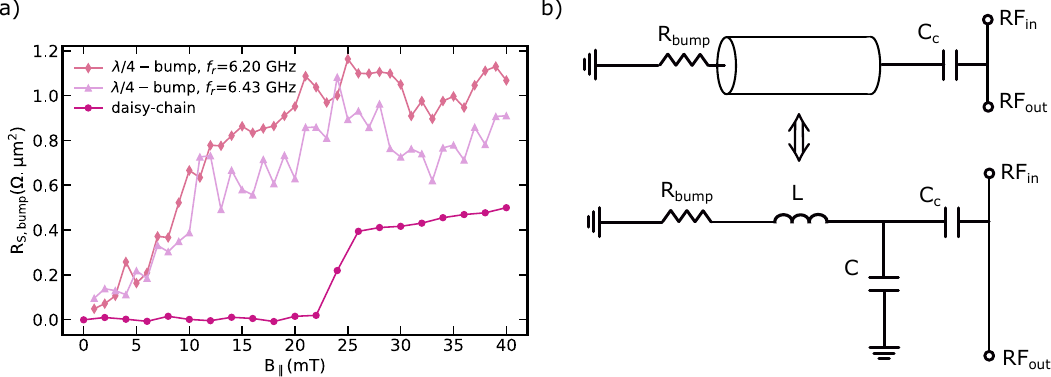}
    \caption{\textbf{Resistance of a bump interconnect:} \textbf{(a)} Bump resistance normalized per µm square area versus in-plane magnetic field $B_{\parallel}$ swept around $B_{c}$(In)$=\SI{23}{\milli\tesla}$. 5-GHz RF resistance of the bump extracted from analysis of the $\lambda /4-$resonator $Q_{i}$ (diamond and triangle) and DC resistance of the bump obtained from a dV/dI-daisy-chain measurement (circle) are reported. \textbf{(b)} Modelling of the type (iii) $\lambda /4-$ resonator by a lossless transmission line loaded to ground with a resistor $R_{bump}$ (top) and equivalent lumped element series $RLC$ circuit model (bottom). The $LC$ resonator is grounded to the top chip with a bump of effective resistance $R_{bump}$ and is $C_c$-capacitively coupled to a transmission line.
     }
     \label{fig:s5}
\end{figure*}

In Fig.{3}{d} of the main text and in Fig.~\ref{fig:s4}{f}, we notice that the internal quality factor of type (iii) resonators decreases continuously to $3 \times 10^4$ in the $5-\SI{25} {mT}$ range before stabilizing at $\mathrm{2 \times 10^4}$. We attribute that decay to the transition of the bump interconnect to the normal state. In order to quantify the effective resistance gained by the bump when it transitions, we model the bump as a series resistance in the lumped-element circuit of the resonator, see Fig.~\ref{fig:s5}{b}. From such a model, the resistor is viewed as a microwave loss source which is responsible for a reduction in $Q_{i}$ of an amount $\mathrm{Q_{bump}=(1/Q_{i}-1/Q_{i, 0})^{-1}}$ where $\mathrm{Q_{i, 0}}$ labels the internal quality factor of the type (iii) $\lambda/4$-resonator measured at $\mathrm{B=\SI{0} {T}}$.
Based on the derivation carried out below, 
the resistance that is related to $\mathrm{Q_{bump}}$ of a $\mathrm{Z_{0}}$ impedance resonator of type (iii) writes: 
\begin{equation}
\mathrm{R_{bump}=\frac{\pi Z_{0}}{4 Q_{bump}}}
\end{equation}
In Fig.~\ref{fig:s5}{a}, we report the surface resistance of a bump ($R_{S,bump}=R_{bump}/S_{bump}$) as a function of the in-plane magnetic field $B_{\parallel}$, with $S_{bump}=\SI{5}{\micro\meter}\times \SI{5}{\micro\meter}$ the area of a resonator bump. We notice around $B_c$(In)$=\SI{23}{\milli\tesla}$ 
Typically, at $\mathrm{B_{\parallel}>\SI{30} {mT}}$, the resistance of the normal state In $\SI{5}{\micro\meter}\times \SI{5}{\micro\meter}$ bump is estimated to be $\mathrm{\SI{50} {m \Omega}}$. We can notice that the normal resistance of the bump extracted from RF measurement is in-line with the DC characterization. Nonetheless, below $B_c(In)$, RF measurements show a higher resistance of the bump compared to the value measured in DC. The origin of this phenomenon is not yet fully understood. \\
In the following, we present the derivation of the expression of the finite resistance $R_{bump}$ of a type (iii) resonator with respect to its internal quality factor $Q_i$.
From \cite{pozar_microwave_2012}, the input impedance of the lossless transmission line resonator of length $l$ loaded to ground with a resistance $R_L=R_{bump}$ is:
\begin{equation}
Z_{in}=Z_{0} \frac{R_L+jZ_0\mathrm{tan}(\beta l)}{Z_0+jR_L\mathrm{tan}(\beta l)}
\end{equation}
Because $l=\lambda/4$ for $\omega=\omega_0$:

\begin{equation}
\beta l=\frac{\pi}{2}+\frac{\pi \Delta \omega}{2\omega_{0}}
\end{equation}

and then:
\begin{equation}
\mathrm{tan}(\beta l)=-\mathrm{cotan}(\frac{\pi \Delta \omega}{2\omega_{0}})\approx-\frac{2\omega_{0}}{\pi \Delta \omega}
\end{equation}
Defining $\delta=\frac{\Delta \omega}{\omega_{0}}$,
the input impedance writes:
\begin{equation}
Z_{in}\approx \frac{Z_{0}R_{L}\pi\delta-2jZ_{0}^2}{Z_{0}\pi\delta - 2jR_L}
\end{equation}
Rearranging the terms of the equation:
\begin{equation}
Z_{in}\approx \frac{Z_{0}^2R_{L}^2\pi^2\delta^2-4Z_{0}^2R_L}{Z_{0}^2\pi^2\delta^2 + 4R_{L}^2}+2j\frac{Z_{0}R_{L}^2\pi\delta-Z_{0}^3\pi\delta}{Z_{0}^2\pi^2\delta^2 + 4R_{L}^2}
\end{equation}

This equation is of the form

\begin{equation}
Z_{in}=R+2j\Delta \omega L
\end{equation}

which is the input impedance of a series RLC resonant circuit. We can identify the resistance of the equivalent circuit as

\begin{equation}
R=\frac{Z_{0}^2R_{L}^2\pi^2\delta^2-4Z_{0}^2R_L}{Z_{0}^2\pi^2\delta^2 + 4R_{L}^2}
\end{equation}

and the inductance of the equivalent circuit as

\begin{equation}
L=\frac{\pi}{\omega_0}\frac{Z_{0}R_{L}^2-Z_{0}^3}{Z_{0}^2\pi^2\delta^2 + 4R_{L}^2}
\end{equation}

The internal quality factor $Q_{i}$ of this resonator can be found as
\begin{equation}
Q_{i}=\frac{\omega_{0}L}{R}
\end{equation}

As a result,
\begin{equation}
Q_{i}=\frac{\pi Z_0(R_{L}^2-Z_{0}^2)}{Z_{0}^2Z_{L}(\pi^2\delta^2-4)}
\end{equation}

Close to resonance $\delta<<1$ and assuming $R_L<<Z_0$, we find:
\begin{equation}
R_L=\frac{Z_{0} \pi}{4Q_{i}}.
\end{equation}

\subsection{$5^\mathrm{th}$-order low-pass filter}
\begin{figure*}[htbp]
     \includegraphics[scale=0.5, angle=90]{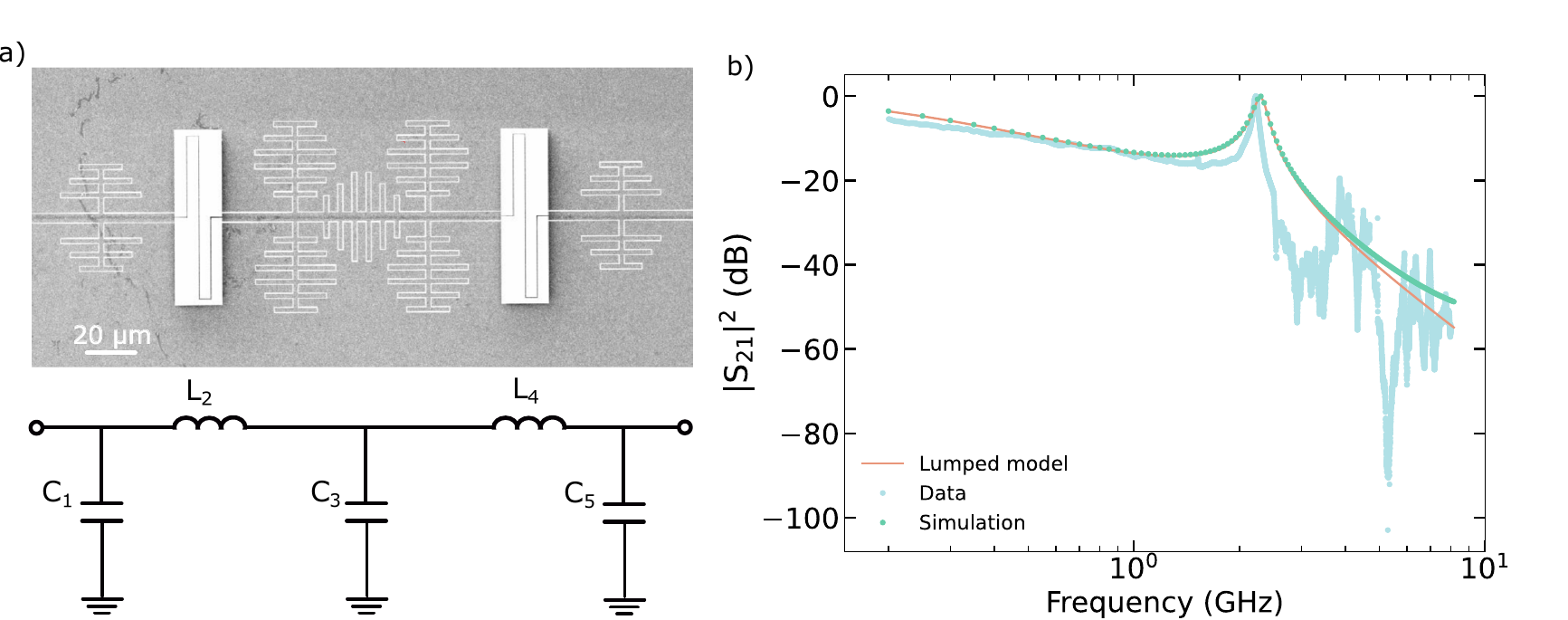}
    \caption{\textbf{$5^\mathrm{th}$-order Butterworth on-chip low-pass filter:} \textbf{(a)} Scanning electron micrograph of $5^\mathrm{th}$-order Butterworth low-pass filter consisting of three fractal interdigital capacitors (C$_{1,3,5}$) and two nanowire inductors (L$_{2,4}$) with is corresponding lumped element circuit below. \textbf{(b)} Measured low-pass filter transmission $|S_{21}|^2$, lumped element prediction and finite element simulation (Sonnet) for a NbN film with $\mathrm{L_{kin}=\SI{130}{\pico\henry}/\square}$.
     }
     \label{fig:s6}
\end{figure*}

In order to mitigate photon loss from the high-impedance resonator to the gates, we present in this study the design of a planar $5^\mathrm{th}$-order low-pass filter. The filter geometry comprises three interdigitate capacitors and two nanowire inductors patterned from a high $\mathrm{L_{kin}}$ NbN thin film, see Fig.~\ref{fig:s6}{a}. \\
We simulated the filter design using Sonnet 2.5D electromagnetic software and fitted in Python the filter response with the ABCD transmission matrix of an appropriate lumped-element circuit model to extract the $\mathrm{C_{k}}$ and $\mathrm{L_{k}}$ design parameters. From the output parameters, it turns out that the devised filter can be accurately described by a lumped-element $5^\mathrm{th}$-order Butterworth low-pass filter in a Cauer topology in the $0-\SI{8} {GHz}$ range, see Fig.~\ref{fig:s6}{b}. In Fig.~\ref{fig:s6}{b}, the transmission $\mathrm{|S_{21}|}$ calculated analytically for an ideal $5^\mathrm{th}$-order Butterworth low-pass filter according to the design parameters is plotted in red, while the transmission $\mathrm{|S_{21}|}$ output from Sonnet electromagnetic simulation of the filter we designed is plotted in green. The simulation and calculation are run for a filter that is loaded with $\SI{50} {\Omega}$ impedance ports. The Sonnet simulation is in good agreement with the lumped-element model up to $\SI{8} {GHz}$. Two regimes can be distinguished with respect to the $\mathrm{f_{c}}=\SI{2} {GHz}$ cut-off frequency of the filter. First, the pass-band domain below $\mathrm{f_{c}}$ is characterized by a ``bending” and a resonance peak at 2 GHz that arise from the filter loading with $\SI{50} {\Omega}$ impedance ports. Above $\mathrm{f_{c}}$, the transmission rolls off sharply with increasing frequency to provide $\SI{-50} {dB}$ attenuation at $\SI{6} {GHz}$. For frequencies that range above $\SI{8} {GHz}$, we observe that the simulation diverges from the circuit model because of a reopening of the transmission spectrum at $\SI{11} {GHz}$ in Sonnet simulations and which originates from parasitic inductances and capacitances that were not taken into account for the model.\\
From a $\mathrm{L_{kin}=\SI{130}{\pico\henry}/\square}$ NbN thin film sputtered on a sapphire substrate, we fabricated $5^{th}$ order low-pass filters embedded into a transmission line and flip-chip to another Sapphire chip covered with the same NbN thin film. The measurement in transmission of the patterned filter is shown in Fig.~\ref{fig:s6}{b} in light blue. We notice that the transmission in the pass-band domain quite accurately fits the behavior expected from the Sonnet simulation and the lumped-element model. (b). The ``bending" and resonance peak at $\SI{2}{GHz}$ arise from the filter loading with $\SI{50}{\Omega}$ impedance ports.
The data shown for $f_{r}>\SI{3}{GHz}$ are the raw measured transmission of the filter to which the on-chip crosstalk has been subtracted. The on-chip cross talk is measured with an open RF circuit consisting of two ports designed on opposite sides of the same chip. The data exhibits a trend similar to the lumped model and the simulation. The sharp roll-off measured at the cut-off frequency, the reopening eye and miscellaneous peaks and dips measured in the 4-$\SI{8}{GHz}$ range are attributed to slot-line modes surrounding the filter which disturb the filter RF measurement.

\subsection{Characterization of the cavity coupled to the DQD}
\begin{figure*}[htbp]
     \includegraphics[scale=0.75, angle=90]{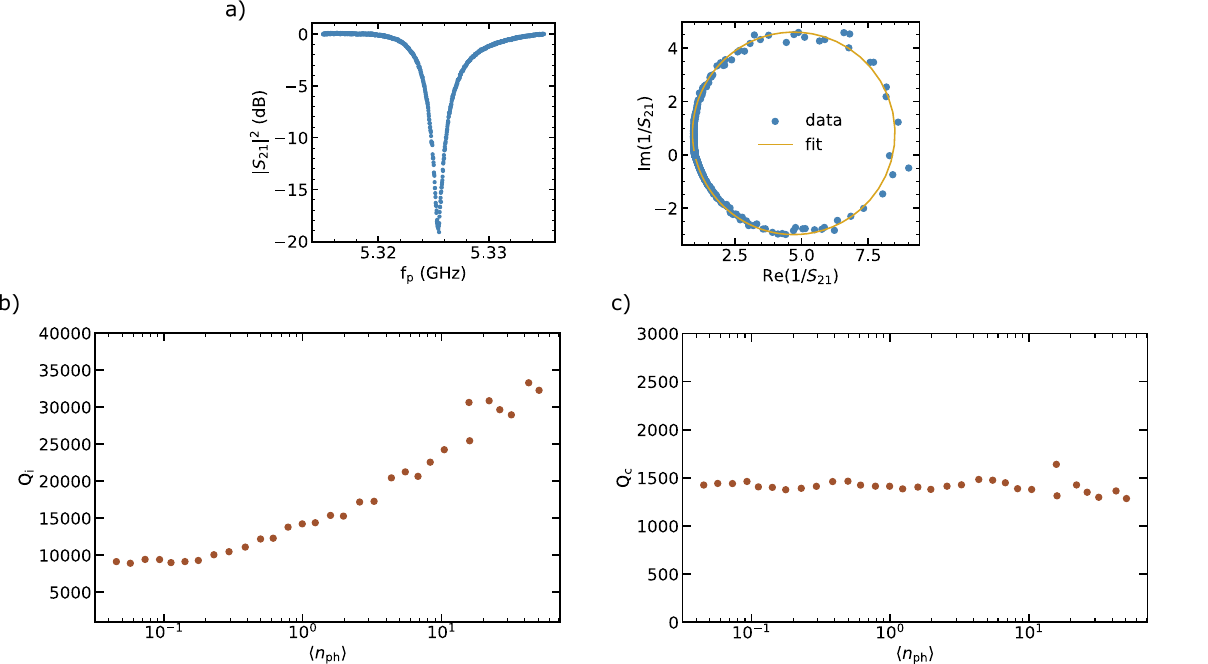}
     \caption{\textbf{Readout cavity characterization:} \textbf{(a)} On the left, transmission $S_{21}$ response of the resonator as a function of the probe frequency $f_{p}$ with $\langle n_{ph} \rangle \approx 0.4$ photon in the resonator. On the right, parametric plot (dot) and fit (solid line) of Im$(1/S_{21})$ vs Re$(1/S_{21})$ of the same data as on the left. $Q_{i}=11100$ is extracted from this fitting procedure. \textbf{(b)} Power dependence of the resonator's internal quality factor. At low photon number, below the single-photon limit, $Q_{i}$ saturates around 10 000. \textbf{(c)} Evolution of the coupling quality factor $Q_c$ with respect to the number of photons in the resonator. $Q_c$ is constant and equal to $Q_c \approx 1500$.}
     \label{fig:s7}
\end{figure*}

Fig.~\ref{fig:s7} reports the characterization of the flip-chip readout cavity at zero magnetic field before forming the QDs. A circle fit of the inverse of the transmission $1/S_{21}$ is performed in the complex plane to extract $Q_{i}$, $ Q_{c}$ and $f_r$, as plotted in Fig.~\ref{fig:s7}{a}. Especially, at $\langle n_{ph} \rangle \approx 0.4$ mean photon number in the cavity, we extract $f_r=\SI{5.325}{\giga\hertz}$, $Q_{i}=11100$ and $Q_{i}=1500$ which correspond to cavity decay rates $\kappa_{i}/2\pi=\SI{480}{\kilo\hertz}$ and $\kappa_{c}/2\pi=\SI{3.64}{\mega\hertz}$. Fig.~\ref{fig:s7}{b} and Fig.~\ref{fig:s7}{c} illustrate the power dependence of $Q_{i}$ and $ Q_{c}$ respectively as a function of the mean number of photons in the resonator.

\subsection{Charge stability diagram}
\begin{figure*}[htbp]
     \includegraphics{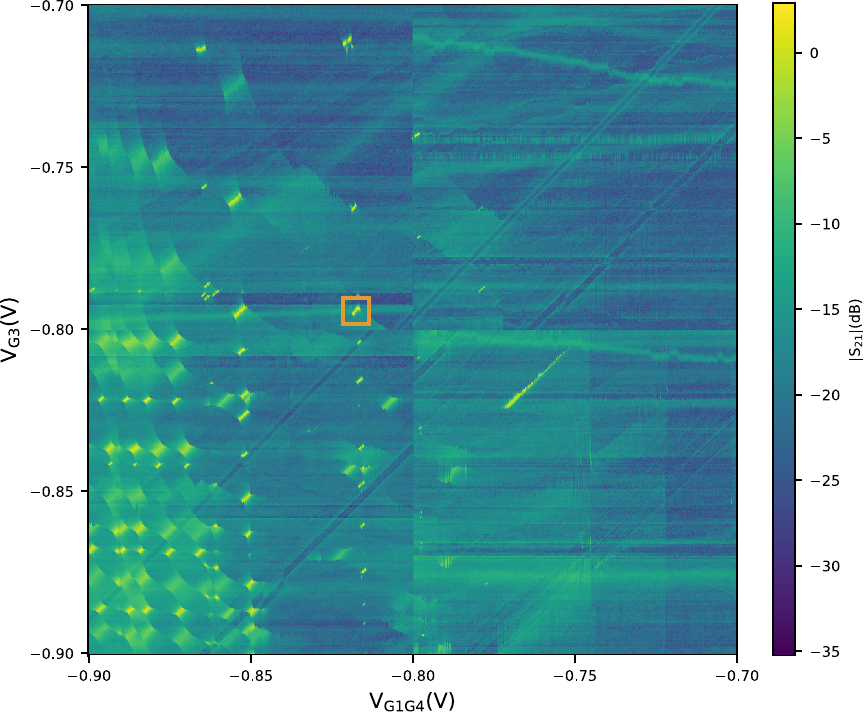}
     \caption{\textbf{Charge stability diagram:} Transmission amplitude probed at the bare cavity frequency $f_{r}=\SI{5.325}{\giga\hertz}$ while sweeping $V_{G3}$ and $V_{G4}$. Four individual measurements are stitched together. The interdot charge transition investigated in this work is highlighted by an orange box.}
     \label{fig:s8}
\end{figure*}

Fig.~\ref{fig:s8} shows the charge stability diagram with respect to the gate voltages $V_{G3}$ and $V_{G4}$ at $V_{SD}=0$V and $V_{G2}=0$V. The transmission is probed at the bare cavity frequency $f_{r}=\SI{5.325}{\giga\hertz}$. As a result, interdot charge transitions are revealed as transmission peaks due to the dispersive downshift of the resonator frequency. The interdot charge transition studied in detail in the main text is highlighted by an orange box. This measurement underlines that the working point is in the few hole regime with $\sim 3$ holes below $G3$ and $\sim 2$ holes below $G4$.

\subsection{Gate lever arm extraction and extraction of charge-photon coupling strength}

\begin{figure*}[htbp]
     \includegraphics[scale=0.85, angle=0]{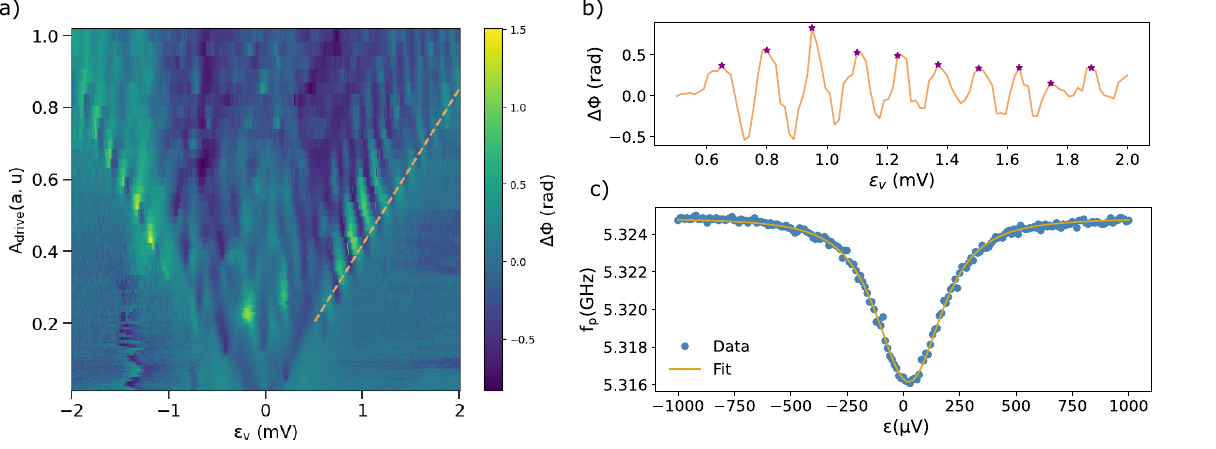}
     \caption{\textbf{Lever arm and charge coupling strength extraction:} \textbf{(a)}Transmission phase measured at the bare resonance frequency as a function of $\varepsilon_{V}$ while applying a \SI{16.21}{GHz} microwave pump tone on $G_4$ at different amplitudes. The phase along the light orange line is plotted in \textbf{(b)} and allows us to evaluate $\alpha=0.546 \pm0.04$. \textbf{(c)} Fitting of the cavity frequency dispersive shift with respect to $\varepsilon_{V}$ and knowing $\alpha$ enables to extract $g_{c}/2\pi=352\pm \SI{13}{\mega\hertz}$ and $t_{c}/h=14.7 \pm \SI{1}{\giga\hertz}$. }
     \label{fig:s9}
\end{figure*}

We use Landau-Zener-Majorana-Stückelberg interferometry to perform spectroscopy of the charge qubit, thereby measuring its frequency $f_{c}$. Then the dependence of $f_{c}$ on voltage detuning $\varepsilon_{V}$ allows us to estimate the lever-arm $\alpha$ \cite{mi_landau-zener_2018}. 
Fig.~\ref{fig:s9}{a} shows transmission through the feedline with respect to voltage detuning while applying a \SI{16.21}{GHz} microwave pump tone on $G_4$ at different amplitudes. As a result, an interference pattern originates from n-photon excitation of the charge qubit when $f_{c}({\varepsilon_{V}})=nf_{pump}$. A linecut across one side of the pattern reveals these multiphoton excitations, see Fig.~\ref{fig:s9}{b}. Due to the interference condition
$\alpha e\delta\varepsilon_{V}=hf_{c}$
where $\delta\varepsilon_{V}$ is the amount of voltage detuning that is required to go from n to n+1 photon excitation of the charge qubit, we extract $\alpha=0.546\pm0.04$.\\
Based on a fit of the dispersive shift of the cavity frequency $f_{r}$ with respect to the voltage detuning $\varepsilon_{V}$, we extract the charge-photon coupling strength $g_{c}$ together with the tunnel coupling rate $t_{c}$ following:
\begin{equation}
\chi_{c}(\varepsilon_{V})=\frac{g_c^{2}d_{c}^{2}}{2\pi}\left( \frac{1}{|f_{c}-f_{r}|}+\frac{1}{f_{c}+f_{r}} \right)
\end{equation}
where 
\begin{equation}
d_{c}=\frac{2t_{c}}{\sqrt{(\alpha \varepsilon_{V}e/h)^2+(2t_{c})^2}}
\end{equation}
is the electric dipole of the charge qubit and $\alpha$ is the detuning lever-arm extracted from Landau-Zener experiment described above.
Fitting the data Fig.~\ref{fig:s9}{c}, we extract $g_{c}/2\pi=352\pm \SI{13}{\mega\hertz}$ and $t_{c}/h=14.7 \pm \SI{1}{\giga\hertz}$. 

\clearpage
\newpage
%


\end{document}